\documentclass[english, superscriptaddress, preprint, aps, amsmath,amssymb,longbibliography]{revtex4-1}
\usepackage[latin9]{inputenc}
\usepackage{graphicx}
\usepackage[usenames,dvipsnames]{xcolor}
\usepackage{amsmath}
\usepackage[resetlabels, labeled]{multibib}
\setcitestyle{super}
\newcites{S}{References Supplementary Materials}

\usepackage{tikz}
\usepackage{tocvsec2}
\usepackage{scrextend}
\usepackage{lineno}
\modulolinenumbers[5]
\makeatletter

\@ifundefined{textcolor}{}
{
 \definecolor{BLACK}{gray}{0}
 \definecolor{WHITE}{gray}{1}
 \definecolor{RED}{rgb}{1,0,0}
 \definecolor{GREEN}{rgb}{0,1,0}
 \definecolor{BLUE}{rgb}{0,0,1}
 \definecolor{CYAN}{cmyk}{1,0,0,0}
 \definecolor{MAGENTA}{cmyk}{0,1,0,0}
 \definecolor{YELLOW}{cmyk}{0,0,1,0}
}

\newcommand{\comment}[1]{}

\usepackage[normalem]{ulem}

\makeatother

\usepackage{babel}

\begin{document}
\title{Robust zero-energy modes \\ in an electronic higher-order topological insulator:\\ the dimerized Kagome lattice}
\author{S. N. Kempkes}
\thanks{Both authors contributed equally.}
\affiliation{Institute for Theoretical Physics, Utrecht University, Netherlands}
\author{M. R. Slot}
\thanks{Both authors contributed equally.}
\affiliation{Debye Institute for Nanomaterials Science, Utrecht University, Netherlands}
\author{J. J. van den Broeke}
\affiliation{Institute for Theoretical Physics, Utrecht University, Netherlands}
\author{P. Capiod}
\affiliation{Debye Institute for Nanomaterials Science, Utrecht University, Netherlands}
\author{W. A. Benalcazar}
\affiliation{Department of Physics, The Pennsylvania State University, University Park, USA}
\author{D. Vanmaekelbergh}
\affiliation{Debye Institute for Nanomaterials Science, Utrecht University, Netherlands}
\author{D. Bercioux}
\affiliation{Donostia International Physics Center, San Sebastian, Spain}
\affiliation{IKERBASQUE, Basque Foundation of Science, Bilbao,  Spain}
\author{I. Swart}
\email{Correspondence to: C.deMoraisSmith@uu.nl, I.Swart@uu.nl}
\affiliation{Debye Institute for Nanomaterials Science, Utrecht University, Netherlands}
\author{C. Morais Smith}
\email{Correspondence to: C.deMoraisSmith@uu.nl, I.Swart@uu.nl}
\affiliation{Institute for Theoretical Physics, Utrecht University, Netherlands}

\date{\today}

\maketitle

\textbf{Quantum simulators are an essential tool for understanding complex quantum materials. Platforms based on ultracold atoms in optical lattices and photonic devices led the field so far, but electronic quantum simulators are proving to be equally relevant. Simulating topological states of matter is one of the holy grails in the field. Here, we experimentally realize a higher-order electronic topological insulator (HOTI). Specifically, we create a dimerized Kagome lattice by manipulating carbon-monoxide (CO) molecules on a Cu(111) surface using a scanning tunneling microscope (STM). We engineer alternating weak and strong bonds to show that a topological state emerges at the corner of the non-trivial configuration, while it is absent in the trivial one. Contrarily to conventional topological insulators (TIs), the topological state has two dimensions less than the bulk, denoting a HOTI. The corner mode is protected by a generalized chiral symmetry, which leads to a particular robustness against perturbations. Our versatile approach to quantum simulation with artificial lattices holds promises of revealing unexpected quantum phases of matter.}

In a visionary colloquium nearly sixty years ago, Feynman proposed to construct so-called quantum simulators - systems that can be engineered and manipulated at will - with the aim of verifying model Hamiltonians and understanding more complex or elusive quantum systems~\cite{Feynman1960,georgescu2014quantum}. It took forty years for the field to properly take off, with the simulation of the Bose-Hubbard model and the superfluid/Mott-insulator transition in a two-dimensional (2D) optical lattice loaded with $^{87}$Rb atoms~\cite{greiner2002quantum}. Since then, triangular, honeycomb, Kagome and other types of optical lattices have been loaded with bosons and/or fermions, and many interesting quantum states of matter have been simulated~\cite{bloch2012quantum}. Later, quantum simulators were realized also in, among others, trapped ion~\cite{blatt2012quantum} and photonic systems~\cite{aspuru2012photonic}. On the other hand, progress on electronic quantum simulators was achieved only very recently. A few years ago, the first artificial electronic lattice was built by positioning CO molecules on a Cu(111) surface, confining the surface-state electrons to a honeycomb lattice~\cite{Manoharan}. The technique was inspired by the pioneering construction of quantum corrals using STM based manipulations of adatoms~\cite{Crommie}. This was followed by other electronic and spin lattices constructed by atomic manipulation in the STM, such as atomic spin chains~\cite{Hirjibehedin,Khajetoorians2011}, the Lieb lattice with $s$-orbitals~\cite{Slot2017, Drost2017} and \textit{p}-orbitals~\cite{Slot2018}, the quasi-crystalline Penrose tiling~\cite{Collins2017}, and the  Sierpi\'{n}ski gasket with a fractional dimension~\cite{Kempkes}.

Besides manipulating the geometry and the dimensionality, it would be desirable to engineer and control \emph{topological properties}~\cite{Moore_2010} in electronic quantum simulators. Topological insulators, superconductors, and semimetals have attracted enormous attention during the last decades, and their potential use in quantum computers has caused a frantic interest in these systems~\cite{NobelHaldane}. In their best known form, TIs are materials that are insulating in the bulk and host topologically protected states in one dimension lower than the bulk~\cite{HasanKane10}. A first example of engineered electronic TIs established by controlled fabrication on the nanoscale is the one-dimensional (1D) Su-Schrieffer-Heeger (SSH) chain~\cite{Drost2017}. 
However, recently it was proposed that another class of topological systems exists, the so-called HOTIs, in which the topological states emerge in at least two dimensions lower than the bulk\cite{BenalcazarScience}. In this way, 0D corner (1D hinge) states were predicted and subsequently observed in a 2D~\cite{noh2018} (3D~\cite{SchindlerBismuth}) TI. At the moment, HOTIs have been experimentally realized in photonic~\cite{noh2018}, phononic~\cite{serragarcia2018}, topolectrical-circuit~\cite{Imhof}, microwave-circuit~\cite{peterson2018}, and acoustic~\cite{xue2018,Ni2018} systems. 
\begin{figure}
    \centering
    \includegraphics[width=0.60\textwidth]{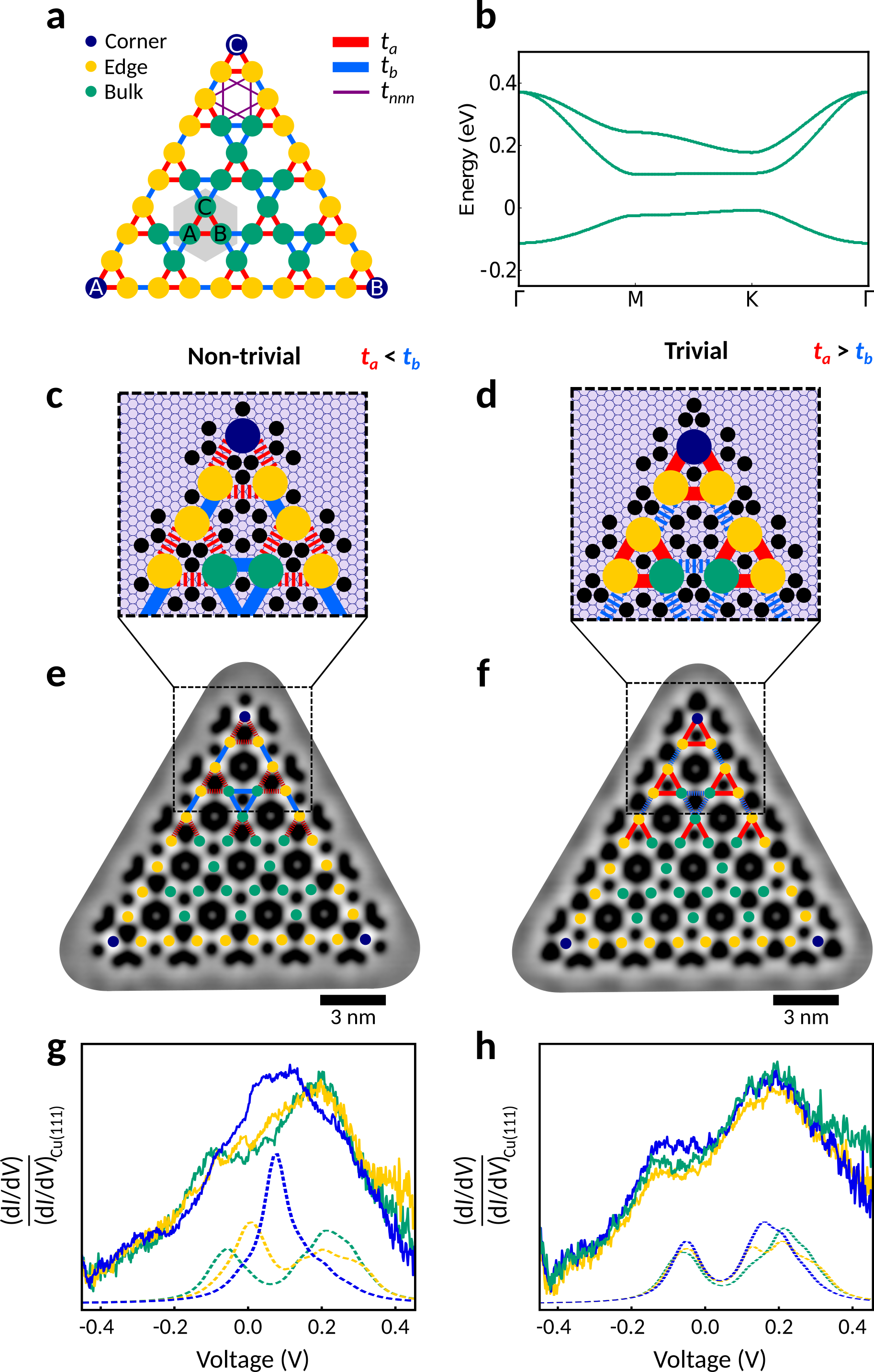}
    \caption{ \textbf{Design of the dimerized Kagome lattice.}  (a) Schematic finite-size tight-binding representation of the Kagome lattice. The unit cell is indicated by a grey hexagon. The model takes both NN hopping ($t_a$ and $t_b$) and NNN hopping ($t_{nnn}$) into account. (b) Band structure for the bulk of the lattice shown in (a), calculated using a tight-binding model with $t_b=75\,$meV, $t_a=0.38 t_b$, $t_{nnn}=0.25 t_b$ and onsite energy $\epsilon=0.075\,$eV. (c-d) Configuration of CO molecules (black) on a Cu(111) surface (grey background) to establish artificial-lattice sites (blue/yellow/green) in a non-trivial ($t_a < t_b$) and trivial ($t_a > t_b$) dimerized Kagome geometry, respectively. Smaller (larger) hopping is indicated by dashed (solid) lines. (e-f) Constant-current STM images of the realized non-trivial and trivial Kagome lattice. Imaging parameters: $I = 0.3\,$nA and $I = 0.1\,$nA, respectively, and $V = 100\,$mV. Scale bars: 3 nm. (g-h) Normalized differential-conductance spectra (solid lines) and the LDOS calculated using the tight-binding model (dashed lines) for the bulk (green), edge (yellow) and corner (blue) sites of the non-trivial and trivial dimerized Kagome lattice, respectively.}
    \label{fig1}
\end{figure}

Here, we present the artificial realization of an \textit{electronic} HOTI. Specifically, we create and characterize a dimerized Kagome lattice~\cite{EzawaPRL}. This lattice, shown in Fig.~\ref{fig1}a, is described by three sites in a unit cell (grey hexagon) with a nearest-neighbor (NN) intracell hopping $t_a$ and intercell hopping $t_b$ (red and blue lines, respectively). The next-nearest-neighbor (NNN) hopping $t_{nnn}$ is indicated in purple only at the top of the lattice (for clarity). In our finite triangular lattice, the corner sites are represented by a blue color, whereas the edge sites are indicated in yellow and the bulk sites in green. The Bloch Hamiltonian (without NNN-hopping) of this model reads
\begin{align}
h_K({\bf k})&=
- \left(
\begin{array}{ccc}
 0 & t_a +t_b e^{i {\bf k \cdot a_2}} & t_a +t_b e^{-i {\bf k \cdot a_3}} \\
 t_a+t_b e^{-i {\bf k \cdot a_2}} & 0 & t_a+t_b e^{-i {\bf k \cdot a_1}} \\
 t_a+t_b e^{i {\bf k \cdot a_3}} & t_a+t_b e^{i {\bf k \cdot a_1}} & 0 \\
\end{array}
\right),
\end{align}
where $\bf k$ is the crystal momentum, and ${\bf a_1}=(1,0)$ and ${\bf a_{2,3}}=(\frac{1}{2},\pm \frac{\sqrt{3}}{2})$ are the lattice vectors. The full tight-binding Hamiltonian that describes the experimentally realized lattice is given in the SI. The bulk band structure is shown in Fig.~\ref{fig1}b. The regular Kagome lattice exhibits a spectrum with a Dirac cone and a flat band. The alternating hopping strengths in the dimerized Kagome lattice $t_a \neq t_b$ open a band gap between the bottom and middle band at the K-point, as displayed for realistic values $t_a = 28.5\,$meV and $t_b = 75\,$meV. Note that the - otherwise flat - top band is dispersive due to a non-negligible NNN hopping $t_{nnn} = 18.8\,$meV (see Methods and SI).

For the finite-size lattice, we distinguish two cases. If $t_a > t_b$, the lattice configuration is topologically trivial and if the values of the hopping amplitudes are switched, \textit{i.e.} $t_a < t_b$, the lattice is topologically non-trivial. In the topological phase, the weakly coupled edge and corner sites are predicted to accommodate edge states and zero-energy corner modes~\cite{EzawaPRL, benalcazar2018charge}, respectively. The edge of the lattice is similar to a one-dimensional SSH model and exhibits gapped bands. In the gap of both the bulk and the edge, three symmetry-protected zero-energy modes arise, which are localized at each of the corners of the lattice.

Usually, the protection of zero-energy topological states is possible in insulators or superconductors that exhibit a symmetric spectrum. In topological superconductors, the particle-hole symmetry enforces this spectral symmetry, and pins the energies of Majorana bound states exactly at zero energy in its Bogoliubov-de Gennes spectrum. In insulators, bipartite lattices provide such spectral symmetry. The bipartite character of a crystal, often known as chiral symmetry, protects an integer number of zero-energy states in 1D systems such as the SSH model (see SI). Recently, it was shown that when additional crystalline symmetries are present, bipartite lattices can also protect zero-energy corner states in 2D HOTIs~\cite{noh2018}. The Kagome lattice, however, is \textit{not} a bipartite lattice, but consists of three sublattices A, B, and C (see unit cell in Fig.~1a). This poses a conundrum because this lattice does exhibit higher-order zero-energy states at $60^\circ$ corners in the topological configuration, despite the absence of the chiral symmetry associated with bipartite lattices. The protection of these zero-energy corner states can be explained by a generalized chiral symmetry, which relies on the fact that the Kagome lattice is tripartite~\cite{Ni2018} (see SI). 


\textit{Lattice realization.} 
Now we turn to the experimental realization of the electronic dimerized Kagome lattice.Figs.~1c-d present the configuration of CO molecules (black) on Cu(111) (grey background) used to constrain the surface-state electrons to the non-trivial and trivial lattice geometry, respectively. Since the CO molecules act as a repulsive barrier to the 2D electron gas at the Cu(111) surface, they are positioned to form the anti-lattice of the Kagome. The distance between the artificial-lattice sites of the Kagome lattice is chosen to be $3\sqrt{3}a \approx 13.3\,$\AA~, where $a \approx 2.56\,$\AA~denotes the Cu(111) NN distance. Strong hopping (solid lines) is established by a wide connection between the sites, while the hopping is weaker (dashed lines) for a narrow connection, implemented by an increased number of CO adsorbates.
The experimental realization of the non-trivial and trivial dimerized Kagome lattice is shown in the constant-current STM images in Figs.~\ref{fig1}e-f. As a guide to the eye, the artificial-lattice sites and the NN hopping are indicated. Differential-conductance spectra were acquired above the bulk (green), edge (yellow) and corner (blue) artificial-lattice sites and normalized by the average spectrum taken on clean Cu(111)~\cite{Manoharan}. We first discuss the spectra acquired above the non-trivial lattice (see Fig.~1g, solid lines). The bulk spectrum (green) shows a peak around a bias voltage of $V = -150\,$mV, which corresponds to the lowest bulk band, and a more pronounced peak around $V =  +200\,$mV, which can be assigned to the middle and top bulk band. The edge spectrum (yellow) exhibits two peaks, located around $V = -20\,$mV and $V = +200\,$mV, indicative of two edge modes. This resembles an SSH chain at the edge with two bands, of which the top band minimum and bottom band maximum are separated by $2(t_b-t_a)$ (without orbital overlap). Around $V = 0\,$mV, a minimum of the bulk and edge spectrum, the corner spectrum (blue) exhibits a maximum. We attribute this peak to a zero-energy mode localized at the corners. In contrast to the non-trivial lattice, the spectra of bulk, edge and corner sites of the trivial lattice are similar. (\textit{cf.} Fig.~1h, solid lines). These results indicate the presence of an electronic zero mode in the non-trivial dimerized Kagome lattice.
The differential-conductance spectra are reproduced by tight-binding calculations of the local density of states (LDOS) at the designated artificial-lattice sites, displayed underneath the experimental spectra in Figs.~1g and 1h (dashed lines), with the same hopping parameters as used in Fig.~1b. The results are further corroborated by muffin-tin calculations (see SI).

\begin{figure}
    \centering
    \includegraphics[width=0.8\textwidth]{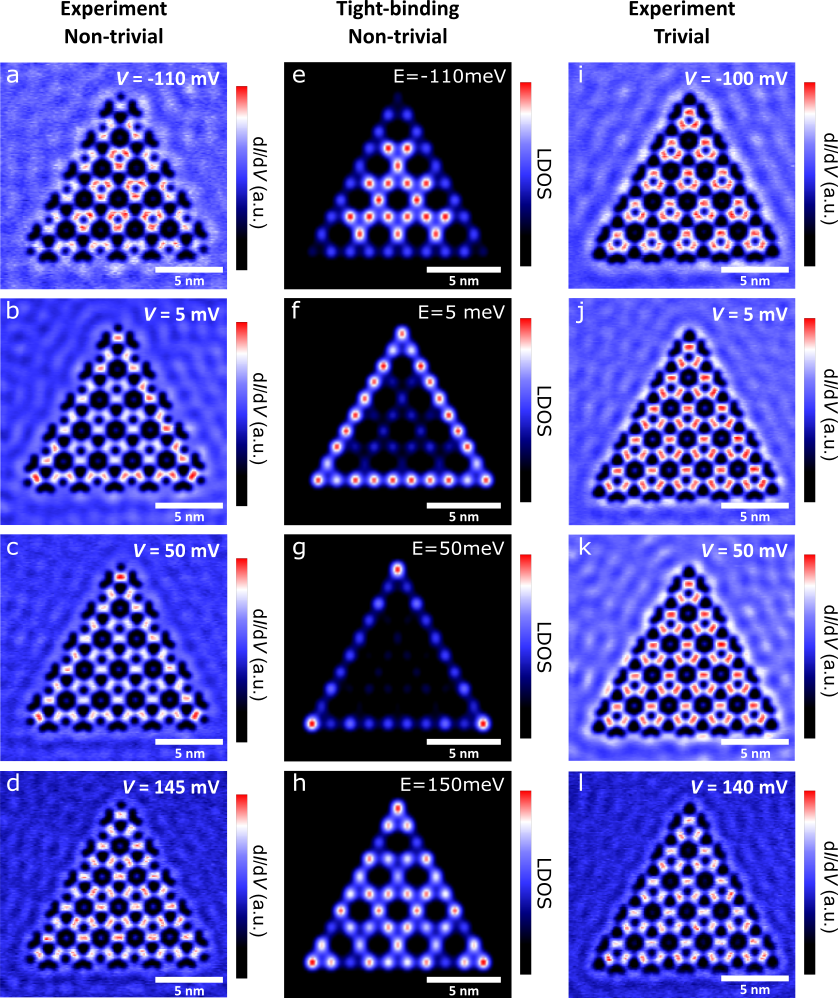}
    \caption{\textbf{Wave-function mapping}. (a-d) Differential-conductance maps acquired above the non-trivial dimerized Kagome lattice at bias voltages $V = -110\,$mV, $V = +5\,$mV, $V = +50\,$mV, and $V = +145\,$mV. (e-h) LDOS maps of the non-trivial lattice at similar energies, simulated using the tight-binding model. (i-l) Differential-conductance maps acquired above the trivial lattice at similar bias voltages.}
    \label{fig2}
\end{figure}

\textit{Wave-function maps.} \\
In Fig.~\ref{fig2}, we investigate the spatial localization of the density of states at bias voltages corresponding to the peak positions in the differential-conductance spectra. Differential-conductance maps acquired above the non-trivial lattice (Figs.~2a-d) are compared with tight-binding calculations (Figs.~2e-h) and differential-conductance maps of the trivial lattice (Figs.~2i-l). At $V = -110\,$mV, the electrons are localized in the bulk of the non-trivial Kagome lattice. Next, at $V = 5\,$meV, the contribution of the bottom edge band is visible. 
At $V = +50\,$mV, we observe the highest intensity at the weakly-connected corner sites, revealing the corner-localized zero modes. Finally, at $V = 145\,$mV, all sites exhibit a similar LDOS, as expected from the spectra. 
These results are in agreement with the tight-binding simulations on the non-trivial lattice (Figs.~2e-h). In contrast, the differential-conductance maps obtained above the trivial lattice show a homogeneous LDOS at all bias voltages. In particular, the corner sites do not exhibit a higher intensity than the other sites at $V = +50\,$mV. The shift of the electron probability from the center to the edges and then finally to the corner sites is only seen for the non-trivial lattice and fully corroborated by tight-binding and muffin-tin calculations (see SI).

\begin{figure}[!h]
    \centering
    \includegraphics[width=1\textwidth]{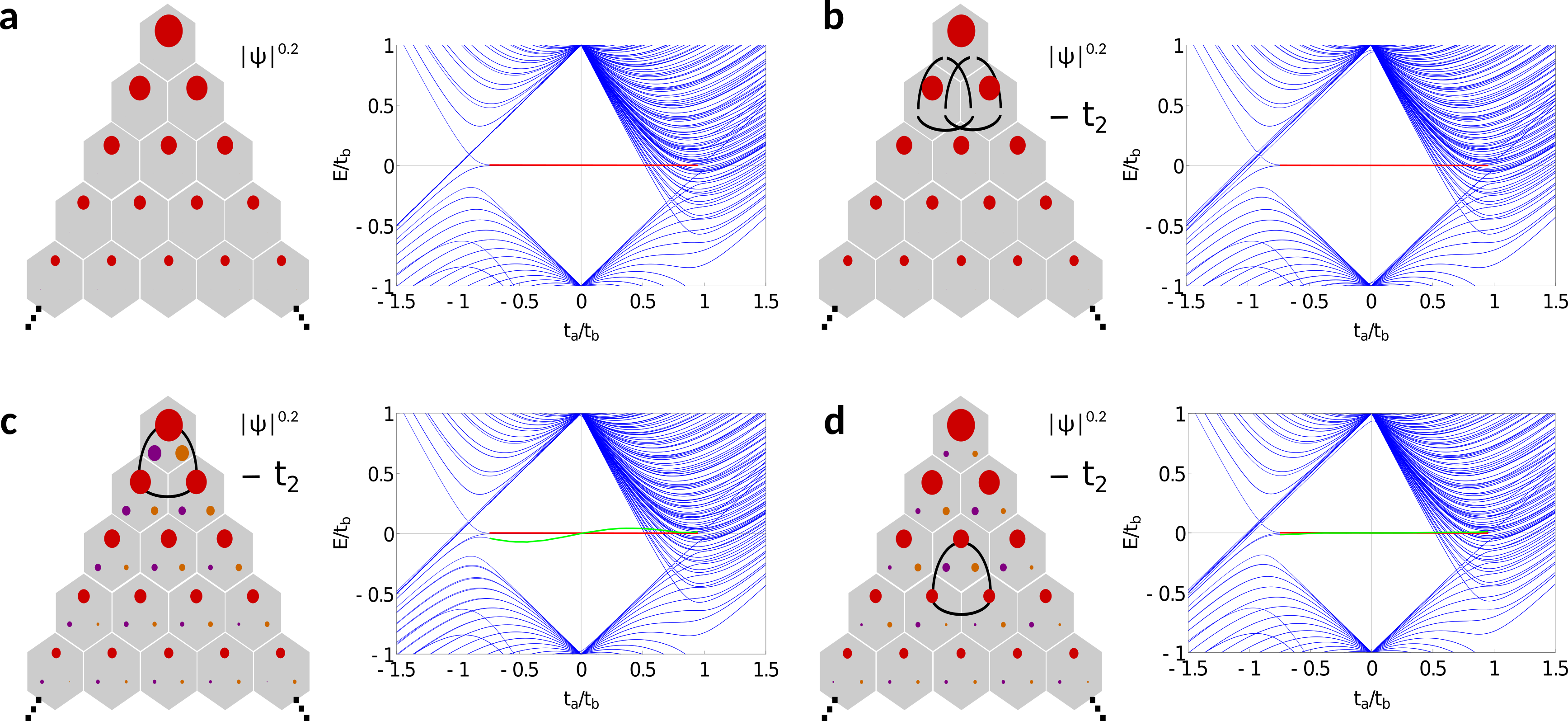}
    \caption{\textbf{Robustness of the zero mode} (a) Localization of the corner mode in the top of a Kagome lattice containing 630 sites. The radius of the circles in the left panel indicates $|\psi|^{0.2}$ to represent the decay of the wave function in a visible way, and the unit cells are indicated with gray hexagons. The corner modes exponentially localize on the corresponding sublattice C (see SI). The spectrum is shown in the right panel, where the zero modes are indicated with a red line. (b) Locally adding a NNN hopping term $t_2=0.05 t_b$ between the A sublattice sites and a similar hopping term between the B sublattice sites breaks the generalized chiral symmetry for the top sites, but this does not affect the zero mode localized at sublattice site C. (c) Breaking the chiral symmetry for the top sublattice site C does shift the zero mode to finite energy and the wavefunction no longer exponentially localizes only on the sublattice sites C. (d) Breaking the chiral symmetry in the bulk also destroys the zero mode and the exponential localization, but the effect of this perturbation is less than in (c). See SI for further analysis on the breaking of these symmetries.}
    \label{fig3b}
\end{figure}
\textit{Zero modes in the Kagome lattice.} \\
The zero modes in the Kagome lattice are protected by the generalized chiral symmetry \cite{Ni2018} (see Methods and SI). To investigate their robustness, we now focus on the top of the lattice, where the corner mode is localized on the sublattice C. Fig.~3a shows that this zero mode has support only on the C sublattice, and decays exponentially in the neighbouring C sites in the bulk and at the edge (the size of the dots represent $|\psi|^{0.2}$ to allow for a visualisation of the decay of the wave function, see SI for the exponential decay of the wave function). If we now locally break the chiral symmetry by introducing a small hopping $t_2=0.05 t_b,$ connecting the A-A and the B-B sites in the neighbourhood of the top corner, the zero mode in C remains unperturbed (see Fig.~3b). On the other hand, if we connect C-C neighbours by a hopping $t_2$, thus locally breaking the chiral symmetry of the C sublattice, the zero mode in C loses protection, moves away from zero energy, and decays also in the A and B sublattices (see Fig. 3c). The other zero modes, at the A and B corners of the lattice, nevertheless, remain unaffected. However, if the local perturbation in C is applied farther away from the corner mode, the disturbance is small (see Fig. 3d). Note that the generalized chiral symmetry is not broken by the NNN hopping $t_{nnn}$ and the orbital overlap, that are present in the experiment (see SI). These results indicate that the generalized chiral symmetry connected to a tripartite system offers more protection to the zero modes than usual bipartite systems do. 

\begin{figure}[!htb]
    \centering
    \includegraphics[width=1.0\textwidth]{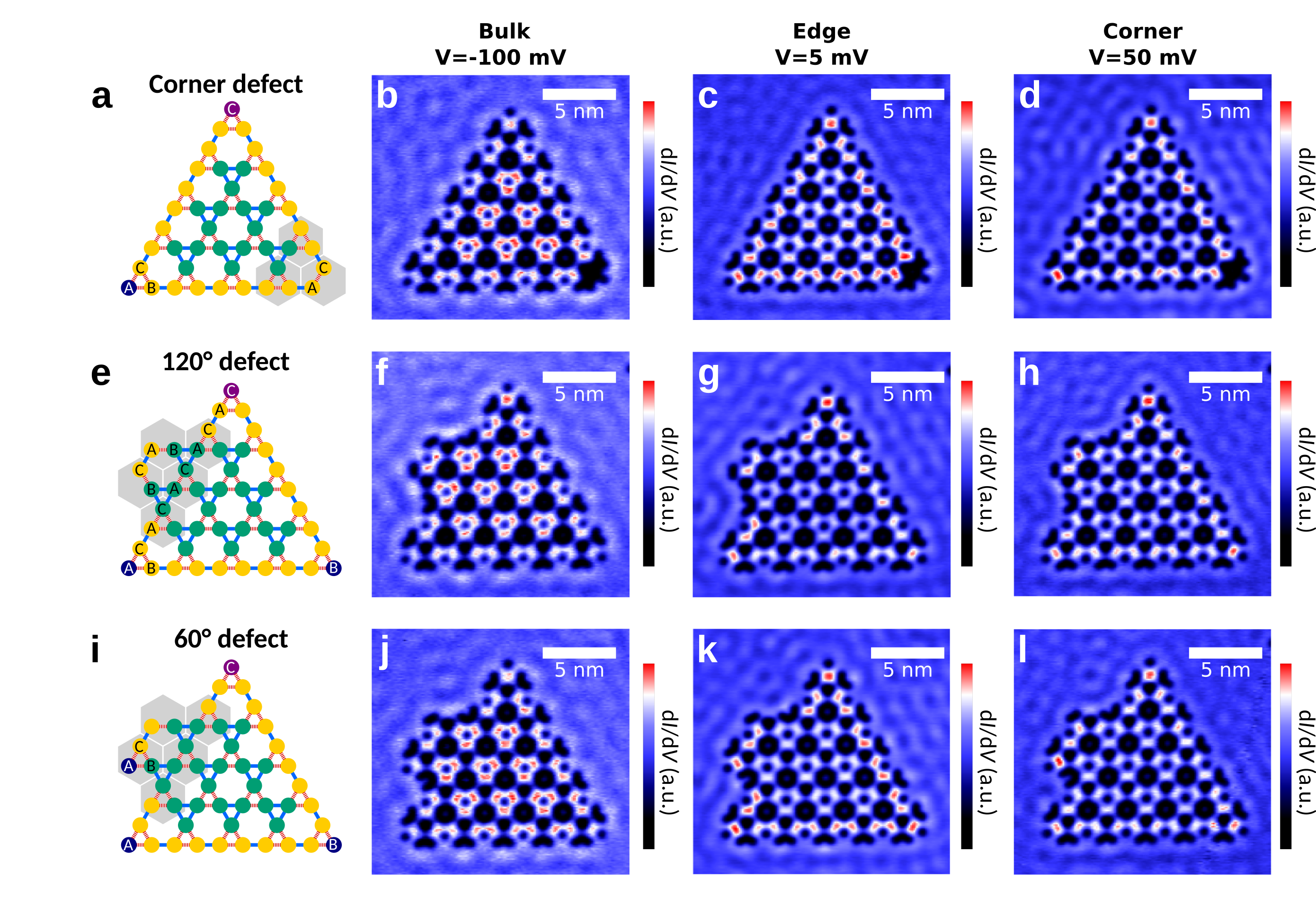}
    \caption{\textbf{Boundary defects in the Kagome lattice.} (a,e,i) Schematics of lattices (a) from which a corner site was removed, (e) with an appendix with two 120$^\circ$ angles added to one edge, and (h) with one 120$^\circ$ and one 60$^\circ$ angle appended. The purple color of the top site in the schematics indicates a slightly lower on-site energy, which is due to an involuntary upward shift of the top CO molecule by $0.256\,$nm. (b-d),(f-h),(j-l) Differential-conductance maps at $V= -100\,$mV (bulk bands), $V= +5\,$mV (edge bands) and $V= +50\,$mV (corner states) for the three lattices.
    }
    \label{fig4}
\end{figure}
\textit{Creating and removing zero modes.} \\
Finally, we show several examples of how these zero modes can also be created and destroyed experimentally by introducing defects into the lattice (see Fig.~\ref{fig4}). In the first defect realization, we remove the corner site at sublattice B (bottom-right corner) from the lattice by blocking the site with CO molecules (see Figs.~\ref{fig4}a-d). Hence, one of the zero modes is no longer present (see Fig.~4a). The corner sites A and C are not affected by the defect, as shown in the differential-conductance map at $V = +50\,$meV in  Fig.~4d. The generalized chiral symmetry is preserved for these modes, as their sublattices remain unperturbed. In this way, by introducing a corner defect we remain with two zero modes.
Second, we append a protrusion at one edge, hosting 120$^\circ$ obtuse angles, breaking the $C_{3}$ symmetry of the lattice but preserving one of its mirror symmetries (see Figs.~4e-h). We observe that the edge mode is disrupted around the positions where the edge no longer consists of only A and C sites (Fig.~4g). On the other hand, the corner modes remain unaltered under this perturbation (see Fig.~4h). This corroborates that, being a local symmetry, the generalized chiral symmetry offers a protection mechanism that is stronger than the one provided by crystalline symmetries: the zero-energy states persist even in the absence of crystalline symmetries.
Finally, a weakly-connected site is added, breaking the mirror symmetry of the entire lattice, but preserving the generalized chiral symmetry. Again, the three zero-energy modes at the corners are resilient. In addition, the added weakly-connected site at $60{^\circ}$ (blue) exhibits a fourth zero-energy mode at sublattice A, protected by the generalized chiral symmetry. Hence, we show that it is possible to create and/or destroy zero-modes at will.

In fact, under the generalized chiral symmetry, zero modes exist whenever a site is only weakly connected to its neighbors (i.e. connected to other sites by hopping terms of amplitude $t_a$, for $t_a<t_b$), as happens in all the cases where zero modes exist in Fig.~\ref{fig4}. Only if two zero modes belonging to different sublattices are in close proximity they can hybridize to open a gap. If, on the other hand, zero modes belonging to the same sublattice are brought together, they will remain at zero energy. In this sense, the generalized chiral symmetry provides a protection mechanism analogous to the conventional chiral symmetry in bipartite lattices, although in this case the existence of three species of zero modes offers more versatility. 

\textit{Conclusion and outlook} \\
The Kagome lattice is known to be a fascinating system, mostly because it realizes geometric frustration and is conjectured to host the elusive spin-liquid phase. Here, we show that a dimerized electronic Kagome lattice brings even further surprises. Protected-zero modes arise at the corners of the lattice, thus realizing a HOTI with extreme robustness due to the tripartite character of the generalized chiral symmetry. By introducing different types of defects into the lattice, zero modes can be manipulated at will, and one can tune the system to have an even or odd number of corner modes. The large degree of control over artificial lattices provides unique opportunities to study electronic topological phases. Using the LDOS as a clear experimental observable, it is possible to detect symmetry-protected edge and corner modes not only at the Fermi energy, but all relevant energies of the model. In addition, the protection mechanisms and robustness of topological phases can be probed by selectively breaking certain symmetries. This can be done either locally, for example via introduction of atomically well-defined defects breaking the crystalline symmetry, or globally, for example by applying a magnetic field. Furthermore, it will be possible to study the influence of disclinations. For topological crystalline insulators, the interplay of topologically protected edge modes and edge geometry can be probed. Electronic quantum simulators are thus complementary to photonic systems, which are designed in a much larger scale, and to the cold-atom setups, which offer great control but request nK temperatures for their operation. The progress in
the realization of artificial electronic structures takes a step forward with the inclusion of topology among the parameters to be manipulated. 


\textbf{METHODS}\\
Methods are available in the online version of the article.

\textbf{Data availability}\\
All data is available from the corresponding authors on reasonable request. The experimental data can be accessed using open-source tools.

\newpage 
\nolinenumbers

\vspace{0.5cm}

\textbf{Supplementary Information} is available in the online version of the article.

\textbf{Acknowledgements} We would like to acknowledge Hans Hansson, Duncan Haldane, and Marcel Franz for fruitful discussions. WAB thanks the support of the Eberly Postdoctoral Fellowship at the Pennsylvania State University. The work of DB is supported by Spanish Ministerio de Ciencia, Innovation y Universidades (MICINN) under the project FIS2017-82804-P, and by the Transnational Common Laboratory $Quantum$---$ChemPhys$. DV, IS and CMS  acknowledge funding from NWO via grants 16PR3245 and DDC13, and DV acknowledges the ERC Advanced Grant 'FIRSTSTEP' 692691 as well. 

\textbf{Author contributions} 
SNK and JvdB performed the theoretical calculations under the supervision of CMS, WAB and DB. MRS, SNK and IS planned the experiment. MRS performed the experiment and data analysis with contributions from PC under the supervision of IS and DV. CMS, SNK and MRS wrote the manuscript with input from all authors.

\newpage

\textbf{METHODS}\\
\textbf{Experiments} \\
The scanning tunneling microscopy and spectroscopy experiments were performed in a Scienta Omicron LT-STM system operating in sample-bias mode at a temperature of $4.5\,$K and a base pressure in the $10^{-11}\,$mbar range. An atomically-flat Cu(111) surface was prepared by several cycles of Ar$^{+}$ sputtering and annealing and was cooled down in the STM head. Carbon monoxide molecules were leaked into the chamber for 20 minutes at a pressure of $1 \cdot 10^{-8}\,$mbar and adsorbed onto the cold surface. The Kagome lattices were assembled and characterized using a Cu-coated platinum-iridium tip, prepared by gentle contact with the Cu(111) surface. CO manipulations were carried out in feedback at $V = 20\,$mV and $I = 40\,$nA, following previously reported procedures\cite{BartelsMethods2, MeyerMethods2, CelottaMethods2}. STM images were obtained in constant-current mode. Differential-conductance spectra and maps were acquired in constant-height mode using a standard lock-in technique with a modulation amplitude of $10\,$mV r.m.s. at a frequency of $273\,$Hz.

\noindent\textbf{Muffin-tin calculations}\\
The muffin-tin model describes the surface-state electrons of the Cu(111) as a 2D electron gas confined between circular potential barriers (CO molecules) with a height of $V=0.9\,$eV and a radius $R=0.3\,$nm. The energy and wave functions for this model can be found by numerically solving the Schr\"odinger equation with this potential landscape. A broadening of $\Gamma =0.08\,$eV is included to account for bulk scattering.

\noindent\textbf{Tight-binding calculations} \\
The free electrons in the lattice act as if they are confined to certain artificial atom positions due to the placing of the CO-molecules. We can describe this behavior within a tight-binding model of connected $s$-orbitals. By making a fit to the experimental and muffin-tin spectra, we are able to determine the hopping parameters, the on-site energy and the orbital overlap. We find the values (in the topological phase) for the strong hopping $t_b=0.075\,$eV and the weak hopping $t_a=0.38 t_b$. Furthermore, we obtain the NNN hopping $t_{nnn}=0.25 t_b$, the on-site energy $\epsilon=0.075\,$eV and the orbital overlap between nearest-neighbours $s_b=0.22$ and $s_a=0.9 s_b$. With these parameters, we solve the generalized eigenvalue equation $H | \psi \rangle = E \mathcal{S} | \psi \rangle$, where $\mathcal{S}$ is the overlap-integral matrix. Next, the LDOS is calculated at each atomic site, in which the broadening $\Gamma=0.08\,$eV is included to account for bulk scattering. Finally, the LDOS maps are calculated by multiplying the LDOS at each site with a Gaussian wave function of width $\sigma = 0.45 d$, where $d = 1.33\,$nm is the distance between two neighboring sites.

\noindent\textbf{Protection mechanism}\\
The protection of the zero modes is due to an extension of the chiral symmetry. The conventional chiral symmetry is expressed as
\begin{align}
\Gamma^{-1} h({\bf k}) \Gamma = -h({\bf k}).
\end{align}
where, without loss of generality, one can choose a basis in which the chiral operator $\Gamma$ is a diagonal matrix with entries $+1$ for one sublattice and of $-1$ for the other one.
In the dimerized Kagome lattice, we have an odd number of lattice sites in the unit cell, and therefore the chiral symmetry does not hold. The concept, however, can be extended to a generalized version of the conventional chiral symmetry because the Kagome lattice is tripartite. The generalized chiral operator, $\Gamma_3$, can now be chosen (by an appropiate choice of ordering in the Hamiltonian matrix) to be a diagonal $3 \times 3$ matrix with entries $\Gamma_3=\text{Diag}(1, e^{2\pi i /3}, e^{-2\pi i /3})$ that differentiates three sublattices~\cite{Ni2018}. The generalized chiral symmetry is then written as
\begin{align}
\Gamma_3^{-1} h_1({\bf k}) \Gamma_3& = h_2({\bf k}), \nonumber\\
\Gamma_3^{-1} h_2({\bf k}) \Gamma_3 &=h_3({\bf k}), \nonumber\\
 h_1({\bf k})+ h_2({\bf k})+ h_3({\bf k})&=0.
\end{align}
In the topological phase, three zero modes exist simultaneously, each of which localizes at one of the three sublattices (see SI). This generalized chiral symmetry does not result in spectral symmetry of the bulk bands. Consequently, bulk bands can also have zero energy, but when the bulk bands are degenerate with the zero modes (for $1/2 < t_a/t_b <1$) they do not mix with the localized corner zero modes. More details on the protecting symmetry and symmetry-breaking perturbations are given in the SI.

\let\oldaddcontentsline\addcontentsline
\renewcommand{\addcontentsline}[3]{}

\end{document}